## On the generation of the quarks through spontaneous symmetry breaking.

**Voicu Dolocan**

Faculty of Physics, University of Bucharest, Bucharest, Romania

**Abstract.** In this paper we present the state of the art about the quarks: group SU(3), Lie algebra, electric charge and the mass. The quark masses are generated in the same way as lepton masses. It is constructed a term in the Lagrangian that couples the Higgs doublet to the fermion fields.

### 1. Quarks

Quark is defined as an elementary particle being a fundamental constituent of matter. The quarks combine to produce composite particles called hadrons, the most stable of which are neutrons and protons, that are the components of atomic nuclei. For example protons consist of two up quarks and one down quark, whereas a neutron is made up of two down quarks and one up quark. Quarks cannot exist independently but as a constituent part of matter. There are six types of quarks: up, down, top, bottom, strange, charm. The antiparticles that correspond to every flavor of quarks are known as antiquarks. Antiquarks have the same mass, same mean lifetime and same spin corresponding to quarks, but their properties like electric charge and other charge have opposite sign [1-7].

Up quarks are the lightest among the quarks. They have the maximum stability due to lowest mass. The symbol used is *u* and its abtiparticle is denoted by $\bar{u}$ , The mass of up quark is ~ 0.39 GeV/c$^2$. Its electric charge is (2/3)*e*.

The down quarks come next to up quarks regarding its light mass. Therefore,it also has high stability. Down quark is denoted by *d* and its antiparticle is denoted by $\bar{d}$ .The mass of down quark is ~ 0.39 GeV/c$^2$. The electric charge is (-1/3)*e*.

The strange quark comes under third lightest among all. Strange quark is denoted by *s* and its antiparticle is denoted by $\bar{s}$ . Its mass is ~0.51 GeV/c$^2$ and its electric charge is (-1/3)*e*.

The charm quark is denoted by *c* and its antiparticle is denoted by $\bar{c}$ . It has the mass ~ 1.6 GeV/c$^2$ and the electric charge (2/3)*e*. The meson which is called as J/Psi particle is an example of charm quark.

The top quark is denoted by *t* and its antiparticle is denoted by $\bar{t}$ . The mass of top quark ia ~ 180 GeV/c$^2$ and its electric charge is (2/3)*e*.

The bottom quark is symbolized by *b* and its antiparticle is denoted by $\bar{b}$ . The mass of bottom quark is ~ 4.8 GeV/c$^2$, and its electric charge is (-1/3)*e*

Quarks were proposed in 1964 by Gell-Mann and Zweig as a means for understanding the SU(3) classification of the hadrons. The hadrons are composite structures built from an elementary triplet of spin ½ quarks, corresponding to the fundamental representation of SU(3). Each quark is assigned spin ½ and baryon number B = 1/3. Baryons are made of three quarks (*qqq*) and the mesons of a quark-antiquark pair $\left(q\bar{q}\right)$. The hypercharge *Y* = *B* + *S* where *S* is the strangeness. The charge is

$$Q = B + \frac{Y}{2} \qquad (1.1)$$

where $T_3$ is the isospin component. The three "flavor" of quarks commonly use up, down and strange, have the properties summarized in Table.1.

Table 1. Properties of the light quarks

| Quark | $T_3$ | S | B | Y | Q |
|---|---|---|---|---|---|
| *u* | ½ | 0 | 1/3 | 1/3 | 2/3 |
| *d* | -1/2 | 0 | 1/3 | 1/3 | -1/3 |
| *s* | 0 | -1 | 1/3 | -2/3 | -1/3 |

The following quark scheme is used

$$p = uud; \quad n = udd; \quad \Delta^{++} = uuu; \quad \Delta^{+} = uud; \; \Delta^{o} = udd; \quad \Lambda^{-} = ddd; \quad \Omega\, sss \qquad (1.2)$$

The *uuu* correctly matches the properties of the doubly charged $\Delta^{++}$-baryon. Its spin J = ½ is obtained by combining three identical J = ½ *u* quarks in their ground state. This completely symmetric ground state *uuu* is forbidden in Fermi statistics. In addition, this naïve quark model is clearly unsatisfactory because there are many possibilities to reproduce the obseved sequence of baryon, antibaryon and meson states. Both problems can be resolved by introducing a new property or quantum number for quarks (no for leptons), "color".We suppose that quarks come in three primary colors, red, green and blue, denoted symbolically by R, G, and B, respectively. We rewrite the quark wave function for the Δ-state in (1.2) as $u_R u_G u_B$ and so we clearly overcome the statistics problems by disposing of the identical quarks. The combination *RGB* is antisymmetric under interchange of a pair of color labels as required by the Fermi statistics of the quarks. The antiquarks are assigned the complementary colors as cyan or antired $\left(\overline{R}\right)$, magenta or antigreen $\left(\overline{G}\right)$ and yellow or antiblue $\left(\overline{B}\right)$. Also, may be written

$$p = RGB; \quad \bar{p} = \bar{R}\bar{G}\bar{B}; \quad \pi = \mathrm{R\bar{R}} + \mathrm{G\bar{G}} + \mathrm{B\bar{B}} \qquad (1.3)$$

**2. The group SU(3), Lie algebra and quarks.**

The set of unitary 3×3 matrices with detU = 1 form the group SU(3). The generators may be taken any $3^2 - 1 = 8$ linearly independent traceless hermitian 3×3 matrices. They are denoted $\lambda_i$ with $i$ = 1,2,… 8. Because SU(2) is a subgroup of SU(3), three of the generators of SU(3) can be constructed from three of SU(2) (Pauli matrices) by extending them to three dimensions. So,

$$\hat{\lambda}_1 = \begin{pmatrix} 0 & 1 & 0 \\ 1 & 0 & 0 \\ 0 & 0 & 0 \end{pmatrix}, \quad \hat{\lambda}_2 = \begin{pmatrix} 0 & -\mathrm{i} & 0 \\ \mathrm{i} & 0 & 0 \\ 0 & 0 & 0 \end{pmatrix}, \quad \hat{\lambda}_3 = \begin{pmatrix} 1 & 0 & 0 \\ 0 & -1 & 0 \\ 0 & 0 & 0 \end{pmatrix} \qquad (2.1)$$

The trace of the hermitian $\lambda_i$ vanishes as required. The commutation relations of the first three generators are similar to those of the 3 Pauli matrices $\sigma_i$ from which they are constructed

$$\left[\hat{\lambda}_i, \hat{\lambda}_j\right]_- = 2i\varepsilon_{ijk}\hat{\lambda}_k, \qquad \mathrm{i,j,k} = \{1,2,3\} \qquad (2.2)$$

$\varepsilon_{ijk}$ is well known totally antisymmetric tensor The remaining five generators can be chosen in different ways

$$\hat{\lambda}_4 = \begin{pmatrix} 0 & 0 & 1 \\ 0 & 0 & 0 \\ 1 & 0 & 0 \end{pmatrix}, \quad \hat{\lambda}_5 = \begin{pmatrix} 0 & 0 & -i \\ 0 & 0 & 0 \\ \mathrm{i} & 0 & 0 \end{pmatrix}, \quad \hat{\lambda}_6 = \begin{pmatrix} 0 & 0 & 0 \\ 0 & 0 & 1 \\ 0 & 1 & 0 \end{pmatrix},$$

$$\hat{\lambda}_7 = \begin{pmatrix} 0 & 0 & 0 \\ 0 & 0 & -\mathrm{i} \\ 0 & \mathrm{i} & 0 \end{pmatrix}, \quad \hat{\lambda}_8 = \frac{1}{\sqrt{3}}\begin{pmatrix} 1 & 0 & 0 \\ 0 & 1 & 0 \\ 0 & 0 & -2 \end{pmatrix} \qquad (2.3)$$

In the definition of $\lambda_8$ the factor $1/\sqrt{3}$ is necessary that the relation

$$tr\hat{\lambda}_i\hat{\lambda}_j = 2\delta_{ij}$$

holds for all $i,j$ = 1,2,…. 8. The $\lambda_3$ and $\lambda_8$ are traceless diagonal matrices, with simultaneous eigenvectors

$$R = \begin{pmatrix} 1 \\ 0 \\ 0 \end{pmatrix}, \quad \mathrm{G} = \begin{pmatrix} 0 \\ 1 \\ 0 \end{pmatrix}, \quad \mathrm{B} = \begin{pmatrix} 0 \\ 0 \\ 1 \end{pmatrix}$$

Relation (2.3) may be rewritten as

$$\left[\hat{\lambda}_i, \hat{\lambda}_j\right]_- = 2if_{ijk}\hat{\lambda}_k \qquad (2.4)$$

where the structure constants are totally antisymmetric under exchange of any two indices

$$f_{ijk} = -f_{jik} = -f_{ikj}, etc \qquad (2.5)$$

The anticommutation relations are as follows

$$\left[\hat{\lambda}_i, \hat{\lambda}_j\right]_+ = \frac{4}{3}\delta_{ij}\mathbf{I} + 2d_{ijk}\hat{\lambda}_k$$
$$d_{ijk} = d_{jik} = d_{ikj}, etc. \qquad (2.6)$$
$$\mathbf{I} = \begin{pmatrix} 1 & 0 & 0 \\ 0 & 1 & 0 \\ 0 & 0 & 1 \end{pmatrix}$$

We have split up the anticommutator (2.6) into a trace term (first term) and a traceless term. It can be verified by means of the explicit representation that

$$tr\hat{\lambda}_i\hat{\lambda}_j = 2\delta_{ij}$$

and accordingly

$$tr\left[\hat{\lambda}_i, \hat{\lambda}_j\right]_+ = 4\delta_{ij} = tr\frac{4}{3}\delta_{ij}\mathbf{I} = \frac{4}{3}\delta_{ij}tr\mathbf{I}$$

because tr(**I**) =3. In Table 2 we present the antisymmetric structure constants $f_{ijk}$ and symmetric coefficients $d_{ijk}$

Table 2. Totally antisymmetric structure constants $f_{ijk}$ and symmetrc coefficients $d_{ijk}$

| ijk | $f_{ljk}$ijk | $d_{ijk}$ | ijk | $f_{ijk}$ijk | $d_{ijk}$ |
|---|---|---|---|---|---|
| 123 | 1 | 3118 1/√3 | 458 | √3/2 | |
| 147 | (1/2) | 146 1/2 | 678 | (√3/2) | 355 1/2 |
| 156 | (-1/2) | 157 1/2 | | | 366 -1/2 |
| 246 | (1/2) | 228 1/√3 | | | 377 -1/2 |
| 257 | (1/2) | 247 -1/2 | | | 448 -1/2√3 |
| 345 | (1/2) | 256 1/2 | | | 558 -1/2√3 |
| 367 | (-1/2) | 338 1/√3 | | | 668 -1/2√3 |
| | | 344 1/2 | | | 778 -1/2√3 |
| | | | | | 888 -1/√3 |

All nonvanishing structure constants are obtained by permutation of the indices listed above. It is useful to redefine the generators as $\hat{F}_i = (1/2)\hat{\lambda}_i$. From (2.4) we deduce that

$$\left[\hat{F}_i.\hat{F}_j\right]_- = if_{ijk}\hat{F}_k \qquad (2.7)$$

Next we introduce the spherical representattion of the *F* operators

$$\hat{T}_\pm = \hat{F}_1 \pm i\hat{F}_2\,, \qquad \hat{T}_3 = \hat{F}_3$$
$$\hat{V}_\pm = \hat{F}_4 + i\hat{F}_5\,, \qquad \hat{\mathrm{Y}} = \frac{2}{\sqrt{3}}\hat{F}_8\,, \qquad \hat{\mathrm{U}}_\pm = \hat{F}_6 \pm i\hat{F}_7 \qquad (2.8)$$

These definitions are based on the observations that the *F* operators are constructed from Pauli matrices. The commutation relations for the new operators are the following

$$\left[\hat{T}_3,\hat{T}_\pm\right]_- = \pm\hat{T}_\pm\,, \qquad \left[\hat{\mathrm{T}}_+,\hat{T}_-\right]_- = 2\hat{T}_3$$
$$\left[\hat{T}_3,\hat{U}_\pm\right]_- = \mp\frac{1}{2}\hat{U}_\pm\,, \qquad \left[\hat{U}_+,\hat{U}_-\right]_- = \frac{3}{2}\hat{Y} - \hat{T}_3 = 2\hat{U}_3$$
$$\left[\hat{T}_3,\hat{V}_\pm\right]_- = \pm\frac{1}{2}\hat{V}_\pm\,, \qquad \left[\hat{V}_+,\hat{V}_-\right]_- = \frac{3}{2}\hat{Y} + \hat{T}_3 = 2\hat{V}_3 \qquad (2.9a)$$
$$\left[\hat{\mathrm{Y}},\hat{\mathrm{T}}_\pm\right]_- = 0\,, \qquad \left[\hat{Y},\hat{U}_\pm\right]_- = \pm\hat{U}_\pm\,, \qquad \left[\hat{Y},\hat{V}_\pm\right]_- = \pm\hat{V}_\pm \qquad (2.9b)$$
$$\left[\hat{T}_+,\hat{V}_+\right]_- = \left[\hat{T}_+,\hat{U}_-\right]_- = \left[\hat{U}_+,\hat{V}_+\right]_- = 0$$
$$\left[\hat{T}_+,\hat{V}_-\right]_- = -\hat{U}_-\,, \qquad \left[\hat{\mathrm{T}}_+,\hat{U}_+\right]_- = \hat{V}_+$$
$$\left[\hat{U}_+,\hat{V}_-\right]_- = \overleftarrow{T}_-\,, \qquad \left[\hat{T}_3,\overline{Y}\right]_- = 0 \qquad (2.9c)$$

We have used the relations resulting from hermicity of $F_i$ , i.e.

$$\hat{T}_+ = (\hat{T}_-)^+ , \quad \hat{U}_+ = (\hat{U}_-)^+ , \quad \hat{V}_+ = (\hat{V}_-)^+ \qquad (2.10)$$

From (2.9) we see that the maximum number of commuting generators of SU(3) Lie algebra is 2 : $[\hat{Y},\hat{T}_\pm]_- = 0$ and $[\hat{T}_3,\hat{Y}]_- = 0$. Therefore SU(3) is of rank 2. We interpret Y as hypercharge and T as isospin, and we define the charge operator as

$$\hat{Q} = \hat{T}_3 + \frac{1}{2}\hat{Y} \qquad (2.11)$$

Further, we discuss the smallest nontrivial representation of SU(3). Due to inherent symmetries of the SU(3) multiplets in the Y-$T_3$ plane we are led to the two equivalent triangles, as shown in Fig.1, which are symmetrically around the origin (Y=0, $T_3$=0). Each representation contains 3 states, which for a particle are denoted by $\Psi_1, \Psi_2, \Psi_3$ (Fig. 1*a*) while for an antiparticle these are denoted by $\overline{\Psi}_1, \overline{\Psi}_2, \overline{\Psi}_3$ (Fig.1*b*). The states $\overline{\Psi}_n$ have opposite hypercharge and opposite $T_3$ component and thus the opposite charge as compared to $\Psi_n$,

$$\hat{Q}\Psi_n = Q_n \Psi_n, \quad \hat{Q}\overline{\Psi}_n = -Q_n \Psi_n \qquad (2.12)$$

Each of the two representations contains an isodoublet T = ½ and an isosinglet T = 0. The isodoublet for a particle is given by the states

$$\Psi_1 = \left|\frac{1}{2}Y\right\rangle, \quad \Psi_2 = \left|-\frac{1}{2}Y\right\rangle \qquad (2.13)$$

whereas the isosinglet is given by the state

$$\Psi_3 = |0Y'\rangle \qquad (2.14)$$

Next we can evaluate the hypercharge, Y. From the eigenvalue equations

$$\hat{T}_3\Psi_1 = +\frac{1}{2}\Psi_1 , \quad T_3\Psi_2 = -\frac{1}{2}\Psi_2 , \quad \hat{T}_3\Psi_3 = 0\Psi_3 \qquad (2.15)$$

can be directly derived the $T_3$ values of $\Psi_1, \Psi_2, \Psi_3$. Considering that $\Psi_1$ is a U-spin singlet, the corresponding value of the hypercharge Y can be derived

$$\hat{U}_3\Psi_1 = 0 \qquad (2.16)$$

From relation (2.9) we have

$$\hat{U}_3 = \frac{1}{4}\left(3\hat{Y} - 2\hat{T}_3\right)$$

so that

$$\hat{Y}\Psi_1 = \frac{1}{3}\left(4\hat{U}_3 + 2T_3\right)\Psi_1 = \frac{2}{3}T_3\Psi_1 = \frac{1}{3}\Psi_1 \qquad (2.17)$$

Since $\Psi_1$ and $\Psi_2$ belong to the same isospin doublet (situated perpendicular to the Y axis) and thus have the same hypercharge, may be written

$$\hat{Y}\Psi_2 = \frac{1}{3}\left(4\hat{U}_3 + 2\hat{T}_3\right)\Psi_2 = \frac{1}{3}\Psi_2 \qquad (2.18)$$

From the relation $[4U_3 + 2\times(-1/2)]\times(1/3) = 1/3$, one obtais $U_3 = ½$. for $\Psi_2$. Accordingly, $\Psi_3$ must have the eigenvalue $U_3 = -1/2$ and thus

$$\hat{Y}\Psi_3 = \frac{1}{3}\left(4\hat{U}_3 + 2\hat{T}_3\right)\Psi_3 = \frac{1}{3}\left[4\times\left(-\frac{1}{2}\right) + 2\times 0\right]\Psi_3 = -\frac{2}{3}\Psi_3 \qquad (2.19)$$

By similar arguments obe obtains for the antiparticle states

$$\hat{Y}\overline{\Psi}_1 = -\frac{1}{3}\overline{\Psi}_1, \quad \hat{Y}\overline{\Psi}_2 = -\frac{1}{3}\overline{\Psi}_2, \quad \hat{Y}\overline{\Psi}_3 = +\frac{2}{3}\hat{\Psi}_3 \qquad (2.20)$$

Therefore, hypercharges are multiplets of one-third. If we accept that the charge operator $\hat{Q}$ is determined by the Gell-Mann –Nishijima relation, we may write

$$\hat{Q}\Psi_1 = \left(\frac{1}{2}\hat{Y} + \hat{T}_3\right)\Psi_1 = \left(\frac{1}{2}\times\frac{1}{3} + \frac{1}{2}\right)\Psi_1 = \frac{2}{3}\Psi_1$$
$$\hat{Q}\Psi_2 = \left(\frac{1}{2}\hat{Y} + \hat{T}_3\right)\Psi_2 = \left(\frac{1}{2}\times\frac{1}{3} - \frac{1}{2}\right)\Psi_2 = -\frac{1}{3}\Psi_2 \qquad (2.21)$$
$$\hat{Q}\Psi_3 = \left(\frac{1}{2}\hat{Y} + \hat{T}_3\right)\Psi_3 = \left(\frac{1}{2}\times\left(-\frac{2}{3}\right) + 0\right)\Psi_3 = -\frac{1}{3}\Psi_3$$

Similarly, for the states of antiparticles one obtains

$$\hat{Q}\Psi_1 = -\frac{2}{3}\Psi_1, \qquad \hat{Q}\Psi_2 = +\frac{1}{3}\Psi_2, \qquad \hat{Q}\Psi_3 = +\frac{1}{3}\Psi_3 \qquad (2.22)$$

Since $\Psi_1$ and $\Psi_2$ form an isodoublet similar to the proton and neutron the $\Psi_1$ quark was named “p quark” and the $\Psi_2$, “n quark”. $\Psi_3$ is named “λ quark”. A more modern nomenclature is “up quark” (*u*), “down quark”(*d)* and “strange quark”. (*s*).

### 3. Quarks masses

The quark masses are generated in the same way as the leptons masses. It is constructed a term in the Lagrangian that couples the Higgs doublet to the fermion fileds

$$L_{\text{fermion mass}} = -G_f\left[\overline{\Psi}_L\phi\Psi_R + \overline{\Psi}_R\overline{\phi}\Psi_L\right] \qquad (3.1)$$

where $\Psi_L$ is left-handed weak-isospin doublet

$$\Psi_L = \begin{pmatrix} u \\ d \end{pmatrix}_L = \frac{1-\gamma^5}{2}\begin{pmatrix} u \\ d \end{pmatrix} \qquad (3.2)$$

with weak hypercharge $Y(q_L) = 1/3$. We specify that now the quark states $\Psi_n$ are denoted by $q_n$. The two right-handed weak-isospin singlets are

$$u_R = \frac{1+\gamma^5}{2}u$$
$$d_R = \frac{1+\gamma^5}{2}d \qquad (3.3)$$

with weak hypercharge $Y(u_R) = 4/3$, $\mathrm{Y}(\mathrm{d}_\mathrm{R}) = -2/3$ and

$$\gamma^5 = \begin{pmatrix} -I & 0 \\ 0 & \mathrm{I} \end{pmatrix}, \qquad \mathrm{I} = \begin{pmatrix} 1 & 0 \\ 0 & 1 \end{pmatrix} \tag{3.4}$$

The fermion Lagrangian (3.1) only gives mass to "down" type fermions, i.e. only to one of the isospin doublet components, where the complex doublet of Higgs boson is as usual

$$\phi = \begin{pmatrix} \phi^+ \\ \phi^o \end{pmatrix} = \frac{1}{\sqrt{2}} \begin{pmatrix} 0 \\ v + H \end{pmatrix} \tag{3.5}$$

with $Y_\phi = +1$. Therefore,

$$\begin{gathered} L_{down} = -G_d \frac{1}{\sqrt{2}} \left[ \begin{pmatrix} \bar{u} & \bar{d} \end{pmatrix}_L \begin{pmatrix} 0 \\ v + H \end{pmatrix} d_R + \bar{d}_R \begin{pmatrix} 0 & \mathrm{v} + \mathrm{H} \end{pmatrix} \begin{pmatrix} u \\ d \end{pmatrix}_L \right] = \\ -\frac{G_d}{\sqrt{2}} (v + H) \left[ \bar{d}_L d_R + \bar{d}_R d_L \right] = -\frac{G_d}{\sqrt{2}} (v + H) \bar{d} d = -m_d \left( 1 + \frac{H}{v} \right) \bar{d} d \end{gathered} \tag{3.6}$$

where $m_d = G_d v/\sqrt{2}$ is the mass of the down quark. To give mass of the *up* quark we need another term in the Lagrangian. It is possible to compose a new term in the Lagrangian , using again a complex (Higgs) doublet in combination with the fermions. The mass Lagrangian for the up type fermions takes the form

$$L_{up} = -G_u \left[ \overline{\Psi}_L \phi_c \Psi_R + \overline{\Psi}_R \phi_c \Psi_L \right] \tag{3.7}$$

where

$$\phi_c = -i \tau_2 \phi^+ = \begin{pmatrix} \bar{\phi}_o \\ -\phi^- \end{pmatrix} = \frac{1}{\sqrt{2}} \begin{pmatrix} v + H \\ 0 \end{pmatrix} \tag{3.8}$$

with $Y_\phi = -1$ and

$$\tau_2 = \begin{pmatrix} 0 & -\mathrm{t} \\ \mathrm{i} & 0 \end{pmatrix}$$

Finally, one obtains

$$\begin{aligned} L_{up} &= -G_u \frac{1}{\sqrt{2}} \left[ \begin{pmatrix} \bar{u} & \bar{d} \end{pmatrix}_L \begin{pmatrix} v+H \\ 0 \end{pmatrix} u_R + \bar{u}_R \begin{pmatrix} v+H & 0 \end{pmatrix} \begin{pmatrix} u \\ d \end{pmatrix}_L \right] = \\ &-\frac{G_u}{\sqrt{2}} (v+H) \bar{u} u = -m_u \left( 1 + \frac{H}{v} \right) \bar{u} u \end{aligned} \qquad (3.9)$$

where $m_u = G_u v/\sqrt{2}$ is the mass of the up quark. The masses depend on the arbitrary couplings $G_{u,d}$ and cannot be predicted. If we take into account the Cabibbo mixing, then suppose the following left-handed quark doublets

$$\Psi_{Lu} = \begin{pmatrix} u \\ d_\theta \end{pmatrix}_L, \quad \Psi_{Lc} = \begin{pmatrix} c \\ s_\theta \end{pmatrix}_L \qquad (3.10)$$

where

$$\begin{aligned} d_\theta &= d\cos\theta_c + s\sin\theta_c \\ s_\theta &= s\cos\theta_c - d\sin\theta_c \end{aligned} \qquad (3.11)$$

where $\theta_c$ is the Cabibbo angle, which describes the orthogonal quark mixing, and has been determined as

$$\cos\theta_c = 0.9737 \pm 0.0025$$

The right-handed weak isospin singlets are

$$R_u = u_R, \quad R_d = d_R, \quad R_c = c_R, \quad R_s = s_R \qquad (3.12)$$

The scalar-fermion interaction Lagrangian is

$$
\begin{aligned}
L_{Yukawa} = &-G_1\left[\left(\bar{u} \quad \bar{d}_\theta\right)_L \begin{pmatrix} v/\sqrt{2} \\ 0 \end{pmatrix} u_R + \bar{u}_R\left(v/\sqrt{2} \quad 0\right)\begin{pmatrix} u \\ d_\theta \end{pmatrix}_L\right] - \\
&G_2\left[\left(\bar{u} \quad \bar{d}_\theta\right)_L \begin{pmatrix} 0 \\ v/\sqrt{2} \end{pmatrix} d_R + \bar{d}_R\left(0 \quad v/\sqrt{2}\right)\begin{pmatrix} u \\ d_\theta \end{pmatrix}_L\right] - \\
&G_3\left[\left(\bar{u} \quad \bar{d}_\theta\right)_L \begin{pmatrix} 0 \\ v/\sqrt{2} \end{pmatrix} s_R + \bar{s}_R\left(0 \quad v/\sqrt{2}\right)\begin{pmatrix} u \\ d_\theta \end{pmatrix}_L\right] - \\
&G_4\left[\left(\bar{c} \quad \bar{s}_\theta\right)_L \begin{pmatrix} v/\sqrt{2} \\ 0 \end{pmatrix} c_R + \bar{c}_R\left(v/\sqrt{2} \quad 0\right)\begin{pmatrix} c \\ s_\theta \end{pmatrix}_L\right] - \\
&G_5\left[\left(\bar{c} \quad \bar{s}_\theta\right)_L \begin{pmatrix} 0 \\ v/\sqrt{2} \end{pmatrix} d_R + \bar{d}_R\left(0 \quad v/\sqrt{2}\right)\begin{pmatrix} c \\ s_\theta \end{pmatrix}_L\right] - \\
&G_6\left[\left(\bar{c} \quad s_\theta\right)_L \begin{pmatrix} 0 \\ v/\sqrt{2} \end{pmatrix} s_R + \bar{s}_R\left(0 \quad v/\sqrt{2}\right)\begin{pmatrix} c \\ s_R \end{pmatrix}_L\right] = I + II + III + IV + V + VI
\end{aligned}
\qquad (3.13)
$$

Further,

$$
I = -G_1 \frac{v}{\sqrt{2}}\left(\bar{u}_L u_H + \bar{u}_R u_L\right) = -G_1 \frac{v}{\sqrt{2}} \bar{u} u = -m_u c^2 \bar{u} u
$$

The mass of the up quark is $m_u c^2 = G_1 v/\sqrt{2}$ .

$$
II = -G_2 \frac{v}{\sqrt{2}}\left(\bar{d}_{\theta L} d_R + \bar{d}_R d_{\theta L}\right) =
$$

$$
-G_2 \frac{v}{\sqrt{2}}\left[\left(\bar{d}_L \cos\theta_c + \bar{s}_L \sin\theta_c\right) d_R + \bar{d}_R\left(d_L \cos\theta_c + s_L \sin\theta_c\right)\right] =
$$

$$
-G_2 \frac{v}{\sqrt{2}} \cos\theta_c\left(\bar{d}_L d_R + \bar{d}_R d_L\right) = -G_2 \frac{v}{\sqrt{2}} \cos\theta_c \bar{d} d = -m_d c^2 \bar{d} d
$$

The mass of the down quark is $m_d c^2 = G_2\left(v/\sqrt{2}\right)\cos\theta_c$

$$III = -G_3 \frac{v}{\sqrt{2}} \left( \bar{d}_{\theta L} s_R + \bar{s}_R d_{\theta L} \right) =$$

$$-G_3 \left[ \left( \bar{d}_L \cos\theta_c + \bar{s}_L \sin\theta_c \right) s_R + \bar{s}_R \left( d_L \cos\theta_c + s_L \sin\theta_c \right) \right] =$$

$$-G_3 \frac{v}{\sqrt{2}} \sin\theta_c \left( \bar{s}_L s_R + \bar{s}_R s_L \right) = -G_3 \frac{v}{\sqrt{2}} \sin\theta_c \bar{s}s = -m_s c^2 \bar{s}s$$

The mass of the strange quark is $m_s c^2 = G_3 \left( v / \sqrt{2} \right) \sin\theta_c$

$$IV = -G_4 \frac{v}{\sqrt{2}} \left( \bar{c}_L c_R + \bar{c}_R c_L \right) = -G_4 \frac{v}{\sqrt{2}} \bar{c}c = -m_c c^2 \bar{c}c$$

The mass of the charm quark is $m_c c^2 = G_4 v / \sqrt{2}$.

$$V = -G_5 \frac{v}{\sqrt{2}} \left( \bar{s}_{\theta L} d_R + \bar{d}_R s_{\theta L} \right) =$$

$$-G_5 \frac{v}{\sqrt{2}} \left[ \left( \bar{s}_L \cos\theta_c - \bar{d}_L \sin\theta_c \right) d_R + \bar{d}_R \left( \bar{s}_L \cos\theta_c - \bar{d}_L \sin\theta_c \right) \right] =$$

$$G_5 \frac{v}{\sqrt{2}} \sin\theta_c \left( \bar{d}_L d_R + \bar{d}_R d_L \right) = G_5 \frac{v}{\sqrt{2}} \sin\theta_c \bar{d}d = -m_d c^2 \bar{d}d$$

$$-G_5 \sin\theta_c = G_2 \cos\theta_c \Rightarrow G_5 = -G_2 \cot an\theta_c$$

$$VI = -G_6 \frac{v}{\sqrt{2}} \cos\theta_c \left( \bar{s}_L s_R + \bar{s}_R s_L \right) = -G_6 \frac{v}{\sqrt{2}} \cos\theta_c \bar{s}s = -m_s c^2 \bar{s}s$$

$$G_6 \cos\theta_c = G_3 \sin\theta_c \Rightarrow G_6 = G_3 \tan\theta_c$$

The Yukawa couplings $G_1$,…..$G_6$ are chosen so that *u, d, s* and *c* are mass eigenstates with the correct quark masses. We have expressed all the mixings in terms of the charge -1/3 quarks, the other conceivable terms were omitted. .

## 4. Conclusions.

We have presented the state of the art about the quarks: group SU(3), Lie algebra, electric charge and the mass [1-7]. The quark masses are generated in the same way as lepton masses. A mass term for a fermion in the Lagrangian would be on the form $-m\bar{\Psi}\Psi$ , but such terms in the Lagrangian are not allowed as they are not gauge invariant. It is constructed a term in the Lagrangian that couples the Higgs doublet to the quarks fields.

**References.**

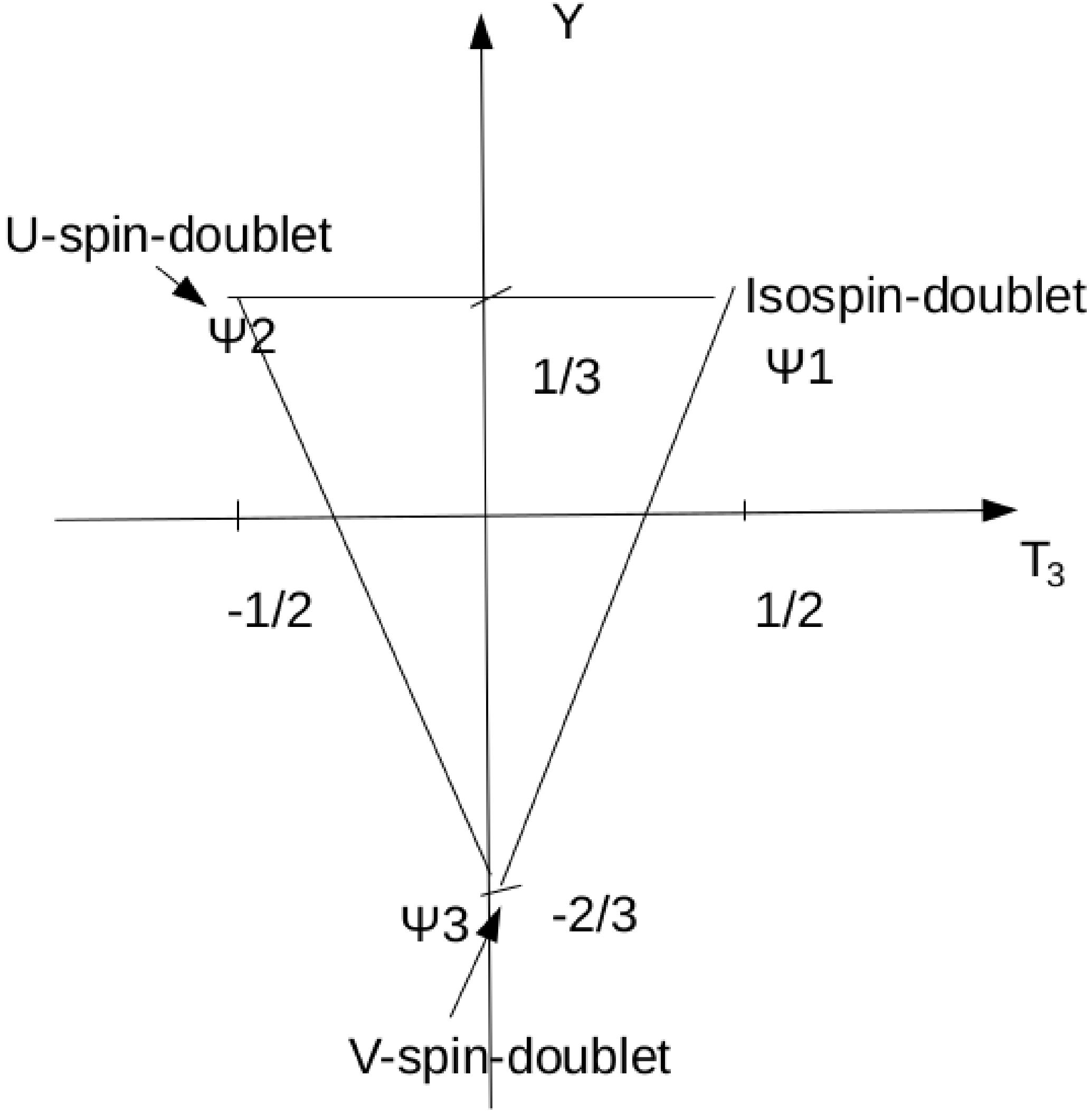

Fig.1*a*. SU(3) quark multiplets: Y = B + S

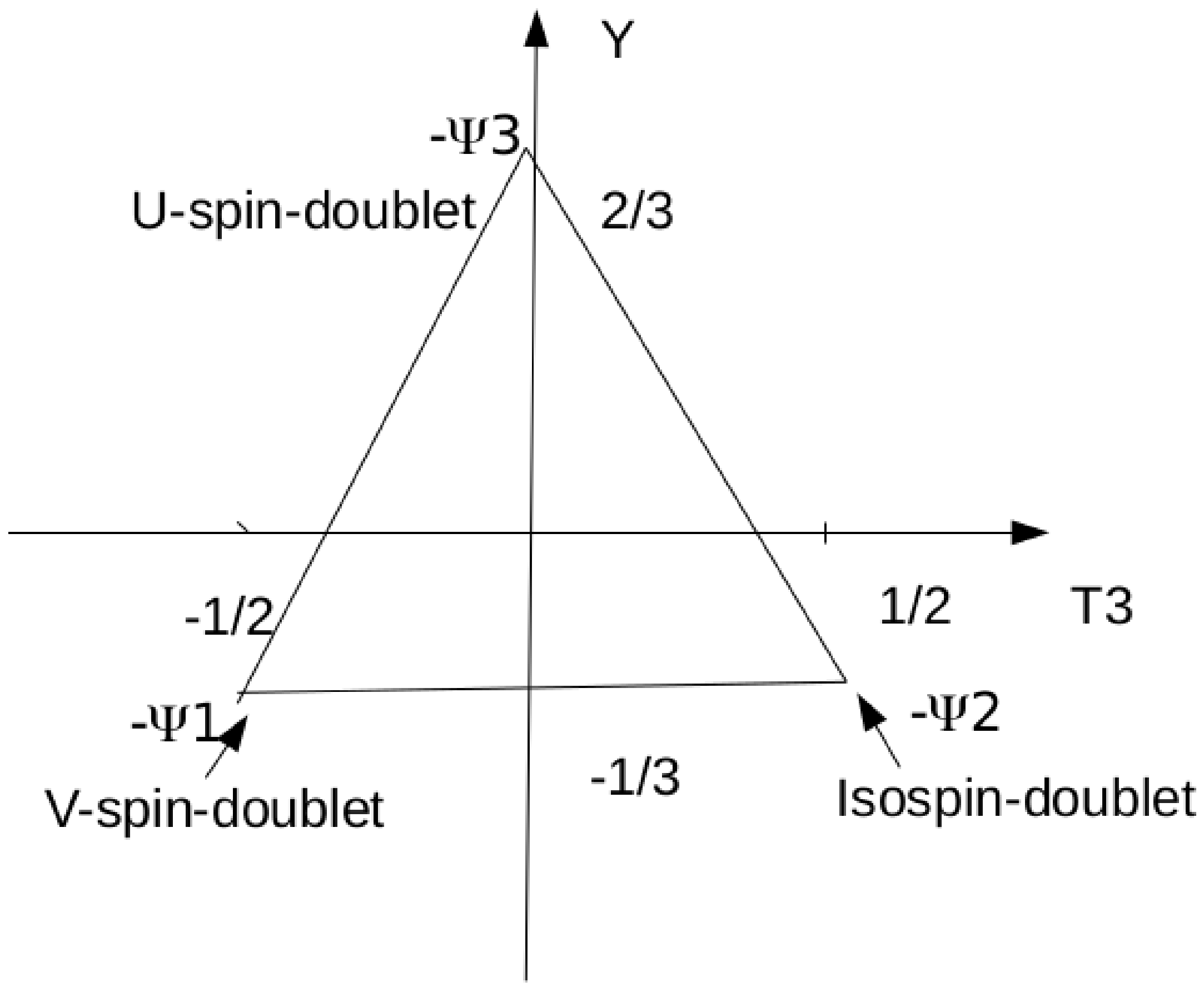

Fig. 1*b*. SU(3) antiquark multiplets: Y = B + S.